\documentclass[preprint,amsmath,aps,10pt,superscriptaddress,letterpaper,tightenlines,showpacs,showkeys]{revtex4}
\usepackage{bm}
\usepackage{stmaryrd}
\usepackage{amsmath}
\usepackage{amssymb}
\usepackage{graphicx}
\usepackage{textcomp}
\usepackage{calrsfs}
\usepackage{yfonts}
\usepackage{color}
\usepackage{amsthm}
\usepackage{lipsum}
\newcommand{\haf}{{\frac{1}{2}}}

\newcommand{\la}{{\langle}}
\newcommand{\ra}{{\rangle}}

\newcommand{\p}{{\partial}}

\begin{document}
\title{A novel derivation of quantum propagator useful for time-dependent trapping and control}

\author{Fardin Kheirandish}
\email{f.kheirandish@uok.ac.ir}
\affiliation{Department of Physics, University of Kurdistan, P.O.Box 66177-15175, Sanandaj, Iran}
\begin{abstract}
\noindent A novel derivation of quantum propagator of a system described by a general quadratic Lagrangian is presented in the framework of Heisenberg equations of motion. The general corresponding density matrix is obtained for a derived quantum harmonic oscillator and a particle confined in a one dimensional Paul trap. Total mean energy, work and absorbed heat, Wigner function and excitation probabilities are found explicitly. The method presented here is based on the Heisenberg representation of position and momentum operators and can be generalized to a system consisting of a set of linearly interacting harmonic oscillators straightforwardly.
\end{abstract}
\maketitle
\section{Introduction}
\noindent The propagators are extensively used in many branches of Physics like quantum electrodynamics, quantum statistical mechanics, condensed matter Physics, polymer Physics, astrophysics and economics \cite{Barton}. They are the main ingredients of theoretical basis of science and engineering. In non-relativistic quantum mechanics, propagators can be calculated in three different approaches \cite{Beau}: (i) Using the explicit form of the energy eigenstates satisfying predefined boundary conditions. (ii) Solving the Heisenberg equations of motion for the problem in question. (iii) Using path integral techniques. The third approach from a pedagogical standpoint is interesting since it points out to the analogy with classical mechanics trough action principle. Due to the formal equivalence of the time-evolution operator in quantum mechanics and the density operator in quantum statistical mechanics, it turns out that the propagator formalism is also useful in calculating partition functions or thermodynamical quantities in quantum statistical mechanics.

In non-relativistic quantum mechanics the propagator gives the probability amplitude for a particle to travel from one spatial point at one time to another spatial point at a later time. This means that, if a system has Hamiltonian $H$, then the appropriate propagator is a function \cite{Sakurai}
\begin{equation}\label{I1}
 G(x,t;x',t')=\frac{1}{i\hbar}\Theta(t-t')K(x,t;x',t'),
\end{equation}
satisfying
\begin{equation}\label{I2}
 (i\hbar\frac{\p}{\p t}-H_x)G(x,t;x',t')=\delta(x-x')\Theta(t-t'),
\end{equation}
where $\Theta(t)$ is the Heaviside step function and $K(x,t;x',t')$ is the kernel of the differential operator in question, often referred to as the propagator instead of $G$ in this context. The propagator can also be written as
\begin{equation}\label{I3}
 K(x,t;x',t')=\la x|\hat{U}(t,t')|x'\ra,
\end{equation}
where $\hat{U}(t,t')$ is the unitary time-evolution operator for the system taking states at time $t$ to states at time $t'$.

The general quadratic Lagrangian Eq.(\ref{GenL}) (see Sec.II), can be transformed to the Lagrangian (\ref{L}) with the corresponding Hamiltonian Eq.(\ref{H}). As a special case, by setting $c(t)=m\omega^2(a-2q\cos(\Omega t))$ and $e(t)=0$, we find the Hamiltonian
\begin{equation}\label{PaulH}
   H=\frac{p^2}{2m}+\haf m\omega^2(a-2q\cos(\Omega t))\, x^2,
\end{equation}
describing the quantum dynamics of a particle in a Paul trap \cite{P1,P2,P3}.

Here we have obtained the quantum propagator for a system described by the general quadratic Hamiltonian Eq.(\ref{H}) which can have applications in quantum control and information processing. A novel derivation is presented in the framework of Heisenberg equations of motion. From the connection between propagator and density matrix, we find the corresponding density matrix for a derived quantum harmonic oscillator and a particle confined in a one dimensional Paul trap. In the following, we will find the total mean energy, work and absorbed heat, Wigner function and excitation probabilities. In the presence of external source ($e(t)\neq 0$), we propose to express the rate of work done by the external source or oscillator and also the absorbed heat by the oscillator, trough expressions that are different from the usual definitions of those quantities in literature. The method can be generalized to a system consisting of a set of linearly interacting harmonic oscillators, straightforwardly.
\section{Lagrangian}
\noindent The general form of a classical quadratic Lagrangian is given by
\begin{equation}\label{GenL}
 L=\haf m \dot{x}^2+a_1 (t)x\dot{x}+\haf a_2 (t)x^2+a_3 (t)\dot{x}+a_4 (t)x,
\end{equation}
regarding the fact that a total time derivative can be removed from Lagrangian, we rewrite the second and fourth terms as
\begin{eqnarray}\label{redef}
  a_1 (t)x\dot{x} &=& \frac{d}{dt}\big(\haf a_1 (t)x^2\big)-\haf \dot{a}_1 (t)x^2,\nonumber\\
  a_3 (t)\dot{x} &=& \frac{d}{dt}\big(a_3 (t)x\big)-\dot{a}_3 (t)x.
\end{eqnarray}
By inserting Eqs.(\ref{redef}) into Eq.(\ref{GenL}) and defining new functions $c(t)=\dot{a}_1 (t)-a_2 (t)$ and $e(t)=\dot{a}_3 (t)-a_4 (t)$, we can rewrite the Lagrangian given in Eq.(\ref{GenL}) as \cite{Schulman}
\begin{equation}\label{L}
 L=\haf m \dot{x}^2-\haf c(t) x^2-e(t) x.
\end{equation}
The conjugate momentum corresponding to the position $x$ is defined by
\begin{equation}\label{momentum}
  p=\frac{\p L}{\p \dot{x}}=m\dot{x},
\end{equation}
and the corresponding Hamiltonian $H=p\dot{x}-L$, is
\begin{equation}\label{H}
  H=\frac{p^2}{2m}+\haf c(t) x^2+e(t) x.
\end{equation}
The system is quantized by promoting classical variables $(x,p)$ to quantum partners $(\hat{x},\hat{p})$ and imposing canonical quantization rule $[\hat{x},\hat{p}]=i\hbar$. From Heisenberg equations for position and momentum, we find
\begin{equation}\label{E1}
 \ddot{\hat{x}}+\frac{c(t)}{m}\hat{x}=-\frac{e(t)}{m}.
\end{equation}
Equation (\ref{E1}) can be solved formally as
\begin{equation}\label{SE1}
  \hat{x}(t)=\hat{x}_H (t)-\frac{1}{m}\int_0^t dt'\,G(t,t')\,e(t'),
\end{equation}
where $G(t,t')$ is the Green's function of Eq.(\ref{E1}) satisfying
\begin{equation}\label{G}
  \bigg[\frac{d^2}{dt^2}+\frac{1}{m}\,c(t)\bigg]G(t,t')=\delta(t-t'),
\end{equation}
and $\hat{x}_H (t)$ is the homogeneous solution
\begin{equation}\label{XH}
  \ddot{\hat{x}}_H (t)+\frac{c(t)}{m}\,\hat{x}_H (t)=0.
\end{equation}
Let $f(t)$ and $g(t)$ be classical independent solutions of Eq.(\ref{XH}), then
\begin{eqnarray}\label{SXH}
&& \ddot{f}(t)+\frac{c(t)}{m}f(t)=0,\nonumber\\
&& \ddot{g}(t)+\frac{c(t)}{m}g(t)=0,\nonumber\\
&& \hat{x}_H (t)=\hat{A} f(t)+\hat{B} g(t),
\end{eqnarray}
where $\hat{A}$ and $\hat{B}$ are operator constants to be determined later. We can find the retarded Green's function in terms of the independent functions $f(t)$ and $g(t)$. The retarded Green's function is defined by
\begin{equation}\label{RG}
  G(t,t')=\left\{
            \begin{array}{ll}
              0, & t<t' ,\\
              af(t)+bg(t), & t>t',
            \end{array}
          \right.
\end{equation}
From the continuity and also the discontinuity of the time derivative of Green's function at $t=t'$, we have
\begin{eqnarray}\label{RG2}
  af(t')+bg(t')=0,\nonumber\\
  \frac{dG}{dt}\bigg|_{t=t^{'+}}-\frac{dG}{dt}\bigg|_{t=t^{'-}}=1,
\end{eqnarray}
by inserting Eq.(\ref{RG}) into Eq.(\ref{RG2}), we find the retarded Green's function as
\begin{equation}\label{FG}
  G(t,t')=\frac{g(t')f(t)-f(t')g(t)}{W(t')}\,\theta(t-t'),
\end{equation}
where $\Theta(t-t')$ is the Heaviside step function and $W(t)$ is the Wronskian defined by
\begin{equation}\label{W}
  W(t)=\det\left(
             \begin{array}{cc}
               g(t) & f(t) \\
               \dot{g}(t) & \dot{f}(t) \\
             \end{array}
           \right).
\end{equation}
From Eqs.(\ref{momentum},\ref{SE1},\ref{FG}), we will find
\begin{eqnarray}
  \hat{x}(0) &=& \hat{x}_H (0)\equiv \hat{x}, \\
  \hat{p}(0) &=& m\dot{x}(0)=m\dot{x}_H (0)\equiv\hat{p}.
\end{eqnarray}
The operator constants $\hat{A},\,\hat{B}$ can be expressed in terms of the initial conditions or Schr\"{o}dinger operators $\hat{x}$, $\hat{p}$ as
\begin{eqnarray}\label{AB}
  \hat{A} &=& \frac{g(0)}{mW(0)}\,\hat{p}-\frac{\dot{g}(0)}{W(0)}\,\hat{x},\nonumber\\
  \hat{B} &=& \frac{\dot{f}(0)}{W(0)}\,\hat{x}-\frac{f(0)}{mW(0)}\,\hat{p}.
\end{eqnarray}
Now by inserting Eqs.(\ref{AB}) into Eq.(\ref{SXH}) and making use of Eq.(\ref{SE1}), we find the position operator as
\begin{equation}\label{SXH2}
  \hat{x}(t) = \frac{g(t)\dot{f}(0)-f(t)\dot{g}(0)}{W(0)}\,\hat{x}+\frac{f(t)g(0)-g(t)f(0)}{W(0)}\,\frac{\hat{p}}{m}-\frac{1}{m}\int_0^t dt'\,G(t,t')e(t').
\end{equation}
\section{Propagator}
In Heisenberg picture, the time evolution of the position operator is given by
\begin{equation}\label{P1}
  \hat{x}(t)=\hat{U}^\dag (t)\hat{x}(0)\hat{U}(t),
\end{equation}
or equivalently
\begin{equation}\label{P2}
  \hat{U}(t)\hat{x}(t)=\hat{x}(0)\hat{U}(t).
\end{equation}
Inserting Eq.(\ref{SXH2}) into Eq.(\ref{P2}) and taking the matrix elements in position space, we find
\begin{equation}\label{P3}
 \la x|U(t)[\alpha(t)x+\frac{\beta(t)}{m}p-\frac{\gamma(t)}{m}]|x'\ra=x\la x|U(t)|x'\ra,
\end{equation}
where
\begin{equation}\label{alfa}
 \alpha(t)=\frac{g(t)\dot{f}(0)-f(t)\dot{g}(0)}{W(0)},
\end{equation}
\begin{equation}\label{beta}
 \beta(t)=\frac{f(t)g(0)-g(t)f(0)}{W(0)},
\end{equation}
\begin{equation}\label{gama}
 \gamma(t)=\int_0^t dt'\,G(t,t')\,e(t').
\end{equation}
From Eq.(\ref{P3}) and the definition $K(x,t;x',0)=\la x|U(t)|x'\ra$, we have
\begin{equation}\label{P4}
  \bigg(x'\alpha(t)+i\hbar\frac{\beta(t)}{m}\frac{\p}{\p x'}-\frac{\gamma(t)}{m}\bigg)K(x,t;x',0)=xK(x,t;x',0),
\end{equation}
where we used $\hat{x}|x'\ra=x'|x'\ra$, $\hat{p}|x'\ra=i\hbar\frac{\p}{\p x'}|x'\ra$. Now Eq.(\ref{P4}) can be rewritten as
\begin{equation}\label{P5}
\frac{\p}{\p x'}\ln K(x,t;x',0)=-\frac{i}{\hbar}\frac{m}{\beta(t)}\,[x-x'\alpha(t)+\frac{\gamma(t)}{m}].
\end{equation}
Since the right side of Eq.(\ref{P5}) is linear in $x'$, we can write
\begin{equation}\label{P6}
  \ln K(x,t;x',0)=C_0 +C_1 x'+\haf C_2 x'^2,
\end{equation}
where the unknown coefficients $C_0$, $C_1$ and $C_2$ may depend on $x$ and $t$. Inserting Eq.(\ref{P6}) into Eq.(\ref{P5}), leads to
\begin{equation}\label{P7}
  (C_1 +C_2 x')=-\frac{im}{\hbar\beta(t)}\,\bigg[x-\alpha(t)\,x'+\frac{\gamma}{m}\bigg],
\end{equation}
and by comparing the coefficients of $x'$ on both sides we obtain
\begin{equation}\label{P8}
  C_1=-\frac{im}{\hbar\beta}\,(x+\frac{\gamma}{m}),\,\,\,\,C_2=\frac{im\alpha}{\hbar\beta}.
\end{equation}
Therefore,
\begin{equation}\label{KK}
  K(x,t;x',0)=e^{C_0 (x,t)}e^{-\frac{im}{\hbar\beta}xx'}e^{\frac{im\alpha}{2\hbar\beta}x'^2}e^{-\frac{i\gamma}{\hbar\beta}x'}.
\end{equation}
Now for the momentum operator we have
\begin{equation}\label{momen1}
 \hat{p}(t)=m\hat{\dot{x}}=m\dot{\alpha}(t)\hat{x}+\dot{\beta}(t)\hat{p}-\dot{\gamma}(t)=U^{\dag}(t)\hat{p}\,U(t),
\end{equation}
so $\hat{U}(t)\hat{p}(t)=\hat{p}\hat{U}(t)$, leading to
\begin{equation}\label{momen2}
 \la x|\hat{U}(t)[m\dot{\alpha}(t)\hat{x}+\dot{\beta}(t)\hat{p}-\dot{\gamma}(t)]|x'\ra=\la x|\hat{p}\,\hat{U}(t)|x'\ra,
\end{equation}
or equivalently
\begin{equation}\label{momen3}
  (m\dot{\alpha} x'+i\hbar\dot{\beta}\frac{\p}{\p x'}-\dot{\gamma})K(x,t;x',0)=-i\hbar\frac{\p}{\p x}K(x,t:x',0).
\end{equation}
By inserting Eq.(\ref{KK}) into Eq.(\ref{momen3}), we find
\begin{equation}\label{momen4}
  \frac{\p C_0 (x,t)}{\p x}=
  \frac{i}{\hbar}[(m\dot{\alpha}+\frac{m}{\beta}-\frac{m\alpha\dot{\beta}}{\beta})x'+\frac{m\dot{\beta}}{\beta}x+
  \frac{\gamma\dot{\beta}-\dot{\gamma}\beta}{\beta}].
\end{equation}
Since the Wronskian does not depend on time ($W(t)=W(0$)), we can easily proof the identity $\alpha\dot{\beta}-\dot{\alpha}\beta=1$, and accordingly, Eq.(\ref{momen4}) can be solved as
\begin{equation}\label{momen5}
  C_0 (x,t)=\frac{im}{2\hbar\beta}\bigg[\dot{\beta} x^2+\frac{2}{m}(\gamma\dot{\beta}-\dot{\gamma}\beta)x\bigg]+\frac{i}{\hbar}\psi(t),
\end{equation}
where $\psi(t)$ is an unknown function of time. The propagator now can be written as
\begin{equation}\label{P9}
  K(x,t;x',0)=e^{\frac{i}{\hbar}\psi(t)}e^{\frac{im}{2\hbar\beta}\big[\dot{\beta}x^2+\alpha x'^2-2xx'+
  \frac{2}{m}(\gamma \dot{\beta}-\dot{\gamma}\beta)x-\frac{2\gamma}{m} x'\big]}.
\end{equation}
By making use of the identity
\begin{equation}\label{unit}
  \la x|\hat{U}(t)\hat{U}^\dag (t)|x'\ra=\la x| \mathbb{I} |x'\ra=\delta(x-x'),
\end{equation}
or equivalently
\begin{equation}\label{unit2}
  \int dx''\, \la x|\hat{U}(t)|x''\ra\la x''|\hat{U}^\dag (t)|x'\ra=\delta(x-x'),
\end{equation}
we find easily
\begin{equation}\label{delta}
 e^{\frac{i}{\hbar}\psi(t)}=\sqrt{\frac{m}{2\pi\hbar\beta}}\,e^{\frac{i}{\hbar}\varphi(t)},
\end{equation}
where $\varphi (t)$ is a real function of time to be determined. Up to now, the quantum propagator has the following form
\begin{equation}\label{P10}
 K(x,t;x',0)= \sqrt{\frac{m}{2\pi\hbar\beta}}\,e^{\frac{i}{\hbar}\varphi(t)}
  e^{\frac{im}{2\hbar\beta}\big[\alpha x'^2+\dot{\beta}x^2-2(\alpha\dot{\beta}-\dot{\alpha}\beta) xx'\big]}\,e^{\frac{i}{\hbar\beta}
  \big[(\gamma \dot{\beta}-\dot{\gamma}\beta)x-\gamma x'\big]}.
\end{equation}
Another well known property of the propagator is
\begin{equation}\label{initial}
  \lim_{t\rightarrow 0} K(x,t;x',0)=\delta(x-x'),
\end{equation}
from Eqs.(\ref{alfa},\ref{beta},\ref{gama}) we have
\begin{eqnarray}
&& \alpha(t)\mathop\rightarrow_{t\rightarrow 0}1,\nonumber \\
&& \gamma(t)\mathop\rightarrow_{t\rightarrow 0}0,\nonumber \\
&& \beta(t)\mathop\rightarrow_{t\rightarrow0} t+o(t^2),\nonumber \\
\end{eqnarray}
where $\beta(t)$ given in Eq.(\ref{beta}), has been expanded by a Taylor series around $t=0$ up to the first order. Note that $\alpha(t)$ and $\gamma(t)$ in Eq.(\ref{P10}) can be simply replaced by their respective limiting values but $\beta(t)$ carrying the singularity, should be replaced with its asymptotic value, namely $t$. Therefore,
\begin{eqnarray}\label{initial2}
 \lim_{t\rightarrow 0} K(x,t;x',0) &=& \lim_{t\rightarrow 0} \frac{e^{\frac{i}{\hbar}\varphi(0)}}{\sqrt{2\pi\hbar t}}\,e^{\displaystyle\frac{i}{2\hbar t}(x-x')^2},\nonumber\\
 &=& \delta(x-x').
\end{eqnarray}
Comparing this result with the well known identity
\begin{equation}\label{delta-func}
  \lim_{t\rightarrow 0} (A/\pi t)^{\frac{1}{2}}e^{-\frac{A}{t}(x-x')^2}=\delta(x-x'),
\end{equation}
we obtain $e^{\frac{i}{\hbar}\varphi(0)}=1/\sqrt{i}$. If we redefine $\lambda(t)=\varphi(t)-\varphi(0)$, then for the quantum propagator we find
\begin{equation}\label{P11}
 K(x,t;x',0)=\sqrt{\frac{m}{2\pi i\hbar\beta}}\,e^{\frac{i}{\hbar}\lambda(t)}
  e^{\frac{im}{2\hbar\beta}\big[\dot{\beta}x^2+\alpha x'^2-2xx'+
  \frac{2}{m}(\gamma \dot{\beta}-\dot{\gamma}\beta)x-\frac{2\gamma}{m} x'\big]}.
\end{equation}
To determine $\lambda(t)$, we know that the propagator $K(x,t;0,0)$ satisfies the Schr\"{o}dinger equation
\begin{equation}\label{Sch1}
  i\hbar\,\p_t K(x,t;0,0)=\bigg[-\frac{\hbar^2}{2m}\p_x^2+\haf c(t)+e(t)x\bigg]K(x,t;0,0),
\end{equation}
taking temporal and spatial differentiations, and setting $x=0$, we find easily
\begin{equation}\label{Sch2}
  \dot{\lambda}(t)=-\frac{1}{2m}\bigg(\frac{\gamma\dot{\beta}-\dot{\gamma}\beta}{\beta}\bigg)^2.
\end{equation}
Finally, the quantum propagator corresponding to the quadratic Lagrangian Eq.(\ref{L}) is
\begin{equation}\label{Finalpro}
  K(x,t;x',0)=\sqrt{\frac{m}{2\pi i\hbar\beta}}\,e^{-\frac{i}{2m\hbar}\int_0^t dt'\,(\frac{\gamma\dot{\beta}-\dot{\gamma}\beta}{\beta})^2}\,e^{\frac{im}{2\hbar\beta}\big[\dot{\beta}x^2+\alpha x'^2-2xx'+
  \frac{2}{m}(\gamma \dot{\beta}-\dot{\gamma}\beta)x-\frac{2\gamma}{m} x'\big]}.
\end{equation}
\section{A differential equation for $\beta(t)$}
\noindent Setting $t'=0$ in Eq.(\ref{FG}) and comparing the result with Eq.(\ref{beta}), we have
\begin{equation}\label{Gbeta}
  G(t)=\beta(t)\,\Theta(t).
\end{equation}
From Eq.(\ref{beta}) it is clear that $\beta(t)$ satisfies the equation
\begin{equation}\label{betaEq}
 \bigg[\frac{d^2}{dt^2}+\frac{1}{m}\,c(t)\bigg]\beta(t)=0,
\end{equation}
with the initial conditions
\begin{equation}\label{Ini-beta}
  \beta(0)=0,\,\,\,\,\frac{d\beta(t)}{dt}\bigg|_{t=0}=1,
\end{equation}
which is the same result obtained by Schulman \cite{Schulman} using a different approach.
\theoremstyle{definition}
\newtheorem{exmp}{Example}
\begin{exmp}
For a harmonic oscillator in an external classical source ($e(t)\neq 0$), we set $c(t)=m\omega^2$ in Eq.(\ref{H}) and find
\begin{eqnarray}
&& f(t)=\cos(\omega t),\nonumber\\
&& g(t)=\sin(\omega t),\nonumber\\
&& \alpha(t)=\cos(\omega t),\nonumber\\
&& \beta(t)=\frac{\sin(\omega t)}{\omega},\nonumber\\
&& G(t,t')=\frac{\sin[\omega(t-t')]}{\omega}\,\theta(t-t'),\nonumber\\
&& \gamma(t)=\int_0^t dt'\,\frac{\sin[\omega(t-t')]}{\omega}\,e(t'),\nonumber
\end{eqnarray}
therefore,
\begin{eqnarray}\label{h}
   K(x,t;x',0) &=& \sqrt\frac{m\omega}{2\pi i\hbar\sin(\omega t)}\,e^{\frac{im\omega}{2\hbar\sin(\omega t)}\big[(x^2+x'^2)\cos(\omega t)-2xx'\big]}\nonumber\\
  && \times\, e^{-\frac{i}{\hbar\sin(\omega t)}\big[x\int_0^t dt'\,\sin(\omega t')e(t')+x'\int_0^t dt'\,\sin\omega (t-t')e(t')\big]}\nonumber\\
  && \times\, e^{-\frac{i}{m\hbar\omega\sin(\omega t)}\int_0^t dt'\int_0^{t'} ds\,e(t')e(s)\sin\omega(t-t')\sin(\omega s)}.\nonumber\\
\end{eqnarray}
\end{exmp}
as expected \cite{Schulman}.
\section{Density matrix}
\noindent The evolution of the density matrix $\rho(t)$ corresponding to a system described by the Hamiltonian Eq.(\ref{H}) is given by
\begin{equation}\label{d1}
  \rho(t)=U(t)\rho (0) U^{\dag}(t),
\end{equation}
where $U(t)$ is the evolution operator connected to the propagator or kernel trough $K(x,t;x',0)=\la x|U(t)|x'\ra$. Therefore, in position basis we can write
\begin{equation}\label{d2}
  \rho(x,x';t)=\int\int_{-\infty}^{\infty}dx_1 dx_2\,K(x,t;x_1,0)\,\rho(x_1,x_2;0)\,K^{*}(x',t;x_2,0).
\end{equation}
By inserting the general form Eq.(\ref{Finalpro}) for the propagator into Eq.(\ref{d2}) we find
\begin{eqnarray}\label{d3}
  \rho(x,x';t) &=& \frac{m\omega}{2\pi\hbar |\beta(t)|}\,e^{\frac{i m}{2\hbar\beta}[\dot{\beta}(x^2-x'^2)+
  \frac{2}{m}(\gamma\dot{\beta}-\dot{\gamma}\beta)(x-x')]}\nonumber \\
   &\times& \int\int dx_1 dx_2\, e^{\frac{i m}{2\hbar\beta}[\alpha x_1^2-2x_1 (x+\frac{\gamma}{m})-\alpha x_2^2 +2 x_2 (x'+\frac{\gamma}{m})]}\,\rho(x_1,x_2;0).
\end{eqnarray}
To proceed, we assume the following gaussian initial state
\begin{equation}\label{d4}
  \rho(x_1,x_2;0)=\sqrt{\frac{\lambda}{\pi}}\,e^{-\frac{\lambda}{2} (x_1^2+x_2^2)},
\end{equation}
leading to
\begin{eqnarray}\label{d5}
  \rho(x,x';t) &=& \sqrt{\frac{\lambda}{\pi \zeta(t)}}\,e^{\frac{i m\alpha(\zeta-1)}{2\hbar\beta\zeta}(x^2-{x'}^2)}\,e^{[\frac{i\alpha\gamma(\zeta-1)}{\hbar\beta\zeta}-\frac{i\dot{\gamma}}{\hbar}](x-x')}\nonumber\\
    &\times & e^{-\frac{\lambda}{2\zeta}[(x+\frac{\gamma}{m})^2+(x'+\frac{\gamma}{m})^2]},
\end{eqnarray}
where for notational simplicity we have defined
\begin{equation}\label{d6}
  \zeta(t)=\frac{1}{\alpha^2 (t)+\frac{\lambda^2\hbar^2\beta^2 (t)}{m^2}}.
\end{equation}
The diagonal elements of the density matrix give the probability density to find the particle at position $x$ at time $t$
\begin{equation}\label{d7}
  P(x,t)=\rho(x,x;t)=\sqrt{\frac{\lambda}{\pi \zeta(t)}}\,e^{-\frac{\lambda}{\zeta(t)}(x+\frac{\gamma}{m})^2}.
\end{equation}
Note that Eq.(\ref{d7}) is a general result obtained for a system governed by the Hamiltonian Eq.(\ref{H}).

For a harmonic oscillator in an external classical field $e(t)$, by setting $\lambda=m\omega/\hbar$ we find $\zeta(t)=1$, and
\begin{eqnarray}\label{d8}
 \rho(x,x';t) &=& \sqrt{\frac{\lambda}{\pi}}\,e^{\frac{-i\dot{\gamma}}{\hbar}(x-x')}\,e^{-\frac{\lambda}{2}[(x+\frac{\gamma}{m})^2+(x'+\frac{\gamma}{m})^2]}.
\end{eqnarray}
If we rewrite Eq.(\ref{d8}) as
\begin{equation}\label{d9}
   \rho(x,x';t)=\la x|\rho |x'\ra=\la x|\psi(t)\ra\la \psi(t)|x'\ra,
\end{equation}
we deduce that the ground state Eq.(\ref{d4}) has evolved to the pure state
\begin{equation}\label{pure}
  \la x|\psi(t)\ra=\big(\frac{\lambda}{\pi}\big)^{1/4}\,e^{\frac{-i\dot{\gamma}}{\hbar}x}\,e^{-\frac{\lambda}{2}(x+\frac{\gamma}{m})^2}.
\end{equation}
The diagonal elements give the position distribution function as
\begin{equation}\label{d10}
  P(x,t)=\sqrt{\frac{m\omega}{\pi\hbar}}\,e^{-\frac{m\omega}{\hbar}(x-a(t))^2},
\end{equation}
where
\begin{equation}\label{d11}
  a(t)=\la x\ra=-\frac{\gamma}{m}=-\frac{1}{m\omega}\,\int_0^t dt'\,\sin[\omega(t-t')]\,e(t'),
\end{equation}
is the position of the center of the Gaussian distribution. The energy of the derived oscillator is defined by
\begin{eqnarray}\label{d12}
  U(t) &=& \mbox{Tr}(\rho H)=\int_{-\infty}^{\infty} dx\,\la x|\rho H|x\ra,\nonumber\\
   &=& \frac{\hbar\omega}{2} +\frac{1}{2}m\omega^2 a^2(t)+\frac{1}{2}m \dot{a}^2 (t)+a(t)\,e(t),
\end{eqnarray}
therefore
\begin{equation}\label{du}
  \frac{d U(t)}{dt}=m\omega^2 a \dot{a}+m\dot{a}\ddot{a}+\dot{a}e+a\dot{e}.
\end{equation}
Let us define the energy of the center of mass as
\begin{equation}\label{center}
  E_c (t)=\frac{1}{2}m\omega^2 a^2(t)+\frac{1}{2}m \dot{a}^2 (t),
\end{equation}
then $U=\hbar\omega/2+E_c +ae$. The probability of finding the derived oscillator in its nth excited state ($|n\ra$) is
\begin{equation}\label{pro}
  P_n (t)=\mbox{Tr}(\rho |n\ra\la n|)=\frac{(\frac{E_c (t)}{\hbar\omega})^n\,e^{-\frac{E_c (t)}{\hbar\omega}}}{n!},
\end{equation}
which is a Poissonian distribution with mean $\la n\ra=E_c (t)/\hbar\omega$.

The power or the rate of work done on the oscillator by external source is
\begin{eqnarray}\label{work}
  \la \frac{dW}{dt}\ra &=& -\la e(t)\,\frac{d\hat{x}}{dt}\ra=-\frac{e(t)}{m}\,\la \hat{p}\ra,\nonumber \\
  &=& -\frac{e(t)}{m}\mbox{Tr}(\rho(t) \hat{p})=-e(t)\,\dot{a}(t),
\end{eqnarray}
which is different from what we obtain from the definition \cite{workdef1,workdef2}
\begin{eqnarray}\label{d14}
  \frac{dW}{dt} &=& -\mbox{Tr}(\rho\frac{\partial H}{\partial t})=-\int_{-\infty}^{\infty} dx\,\la x|\rho\frac{\partial H}{\partial t}|x\ra,\nonumber\\
   &=& -a(t)\dot{e}(t).
\end{eqnarray}
From Eqs.(\ref{du},\ref{d14}) we recover the thermodynamic relation
\begin{equation}\label{d15}
  \frac{dU}{dt}=\frac{dQ}{dt}-\frac{dW_{osc}}{dt},
\end{equation}
if we define
\begin{equation}\label{heat}
  \frac{dQ}{dt}=m\omega^2 a(t)\dot{a}(t)+m\dot{a}(t)\ddot{a}(t)+2\dot{a}(t)\,e(t)+a\dot{e},
\end{equation}
which is again different from the definition \cite{workdef1,workdef2}
\begin{eqnarray}\label{d13}
  \frac{dQ}{dt} &=& \mbox{Tr}(\frac{\partial\rho}{\partial t}\,H)=\int_{-\infty}^{\infty} dx\,\la x|\frac{\partial\rho}{\partial t}\,H|x\ra,\nonumber\\
   &=& m\omega^2 a(t)\dot{a}(t)+m\dot{a}(t)\ddot{a}(t)+\dot{a}(t)\,e(t),
\end{eqnarray}
if there is a non zero external source ($e(t)\neq 0$). Note that from the definition of energy Eq.(\ref{d12}) we have
\begin{equation}\label{new1}
 \frac{dU}{dt}=\mbox{Tr}(\frac{\partial\rho}{\partial t} H)+\mbox{Tr}(\rho \frac{\partial H}{\partial t}),
\end{equation}
which can be rewritten as
\begin{equation}\label{new2}
  \frac{dU}{dt}=\bigg(\mbox{Tr}(\frac{\partial\rho}{\partial t} H)+\chi(t)\bigg)+\bigg(\mbox{Tr}(\rho \frac{\partial H}{\partial t})-\chi(t)\bigg),
\end{equation}
where $\chi(t)$ is a correction term given by
\begin{equation}\label{new3}
  \chi(t)=\dot{a}e+a\dot{e}=\frac{d(a(t) e(t))}{dt}.
\end{equation}
The term $a(t) e(t)$ can be interpreted as the energy of the center of mass of the Gaussian distribution in the external field. Therefore, we may define the rate of absorbed energy and work the work done by the oscillator respectively as
\begin{eqnarray}\label{new4}
 \frac{dW}{dt} &=& \mbox{Tr}\bigg(\frac{\partial\rho}{\partial t} H\bigg)+\chi(t)),\nonumber\\
 \frac{dQ}{dt} &=& -\mbox{Tr}\bigg(\rho \frac{\partial H}{\partial t}\bigg)+\chi(t).
\end{eqnarray}

For a particle trapped in a Paul trap, we set $c(t)=m\omega^2[a-2q\cos(\Omega t)],\,e(t)=0,\,\lambda=m\omega/\hbar$, and find the density matrix as
\begin{eqnarray}\label{trapden}
  \rho(x,x';t) &=& \sqrt{\frac{\lambda}{\pi \zeta(t)}}\,e^{\frac{i m\alpha(\zeta-1)}{2\hbar\beta\zeta}(x^2-{x'}^2)}\,e^{-\frac{\lambda}{2\zeta}(x^2+{x'}^2)},\nonumber\\
  &=& \la x|\varphi(t)\ra\la \varphi(t)|x'\ra.
\end{eqnarray}
The corresponding Wigner function \cite{Wigner} can also be determined from the Weyl symbol of the density matrix
\begin{eqnarray}\label{Wigner}
  W(x,p) &=& \int_{-\infty}^\infty dy\,\la x-y/2|\hat{\rho}|x+y/2\ra\,e^{\frac{i}{\hbar}yp},\nonumber\\
         &=& 2\,e^{-\frac{\lambda}{\zeta}x^2}e^{-\frac{\zeta}{\lambda\hbar^2}(p-\frac{m x\alpha(\zeta-1)}{\beta\zeta})^2}.
\end{eqnarray}
From Eq.(\ref{trapden}) it is seen that the initial state (\ref{d4}) has evolved to the final state
\begin{equation}\label{trapstate}
 \varphi(x,t)=\bigg(\frac{m\omega}{\pi\hbar\zeta}\bigg)^{1/4}\,e^{-\frac{m\omega}{2\hbar\zeta}[1-\frac{i\alpha(\zeta-1)}{\omega\beta}]\,x^2}.
\end{equation}
The position distribution function is obtained by setting $x=x'$, in density matrix or from $\int dp/2\pi\hbar\,W(x,p)$ as
\begin{equation}\label{d16}
  P(x,t)=\sqrt{\frac{m\omega}{\pi\hbar\,\zeta(t)}}\,e^{-\frac{m\omega}{\hbar\zeta(t)}x^2}.
\end{equation}
To find $\zeta(t)$, we insert $c(t)=m\omega^2[a-2q\cos(\Omega t)]$ into Eq.(\ref{SXH}) and note that in this case the independent functions $f(t)$ and $g(t)$ satisfy the Mathieu equation \cite{NIST}
\begin{equation}\label{d17}
  \ddot{y}(t)+\omega^2[a-2q\cos(\Omega t)]\,y(t)=0.
\end{equation}
In terms of the dimensionless variable $u=\omega t$, the Mathieu equation is transformed to the standard form
\begin{equation}\label{d17n}
  \frac{d^2 y(u)}{du^2}+[a-2q\cos(2ru)]\,y(u)=0,
\end{equation}
where $2r=\Omega/\omega$. The equation (\ref{d17n}) can be solved using the \emph{Mathemmatica} software, and the independent solutions are given by
\begin{eqnarray}\label{d18}
  f(u) &=& \mbox{MathieuC}[a/r^2,q/r^2,ru],\nonumber \\
  g(u) &=& \mbox{MathieuS}[a/r^2,q/r^2,ru].
\end{eqnarray}
where $\mbox{MathieuC}$ and $\mbox{MathieuS}$ are Mathieu functions. Knowing $f(u)$ and $g(u)$, we can determine the functions $\alpha(u)$ and $\beta(u)$ from Eqs.(\ref{W},\ref{alfa},\ref{beta}), note that $\alpha(t)=\alpha(u)$ but $\omega^2\beta(t)=\beta(u)$. Finally, $\zeta(u)$ can be determined from Eq.(\ref{d6}) as
\begin{equation}\label{explicit}
  \zeta(u)=\frac{1}{\alpha^2(u)+\beta^2(u)},
\end{equation}
where we have assumed $\lambda=m\omega/\hbar$. The behaviour of $\zeta(u)$ is depicted in Fig.1 for the values $a=1,\,r=10,\,q=0.25$.
\begin{figure}[ht]
\centering
\includegraphics[width=0.45\textwidth]{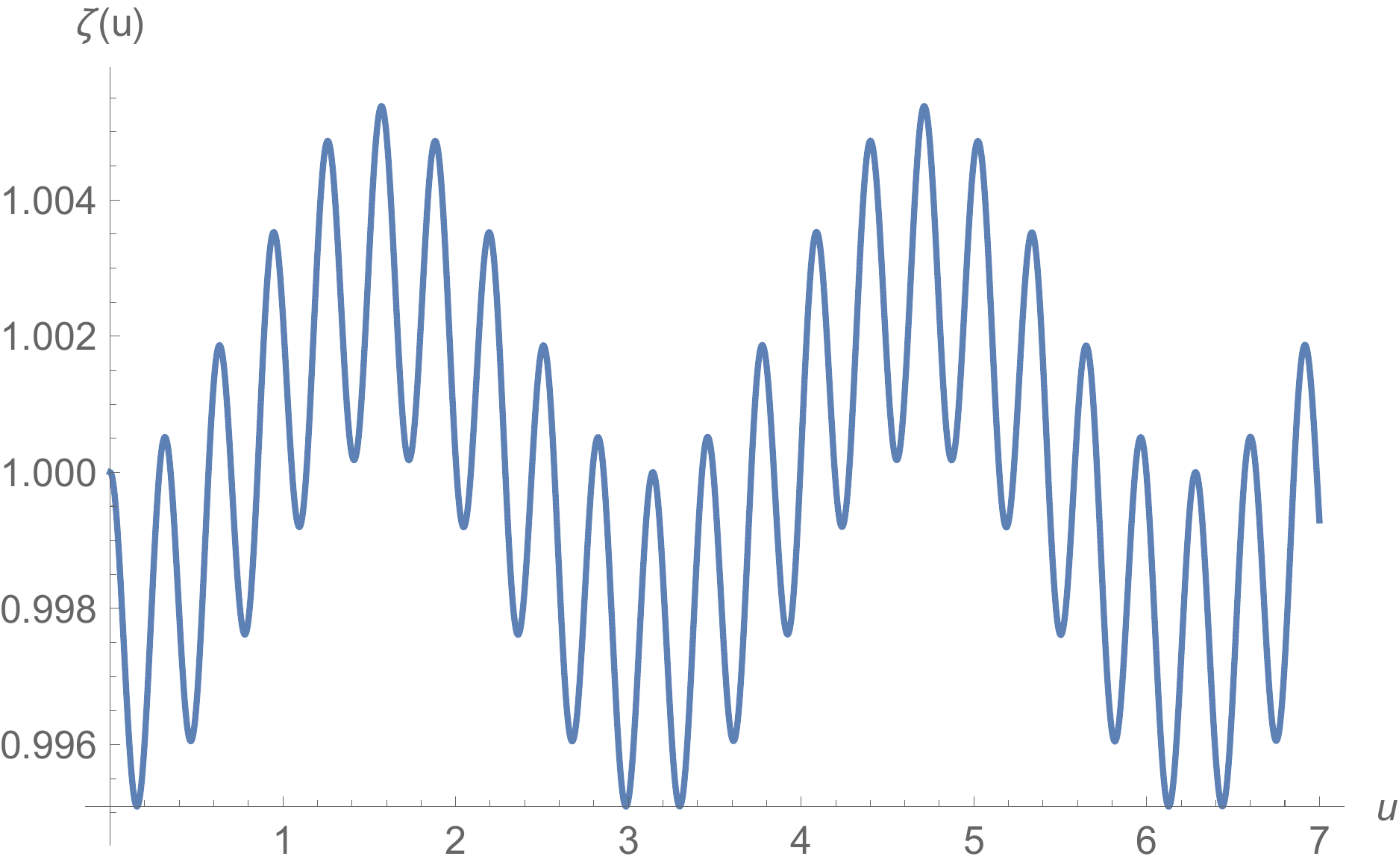}
\caption{The behaviour of $\zeta(u)$ for values $a=1,\,r=10,\,q=0.25 $. The minimum of $\zeta(u)$ is strictly greater than zero due to the uncertainty principle ($\Delta x=\sqrt{\hbar\zeta(u)/2m\omega}>0$). There is a squeezing in $\Delta x$, ($\Delta x<\sqrt{\hbar/2m\omega}$) in intervals where $\zeta(u)<1$.}
\end{figure}

\begin{figure}[ht]
\centering
\includegraphics[width=0.45\textwidth]{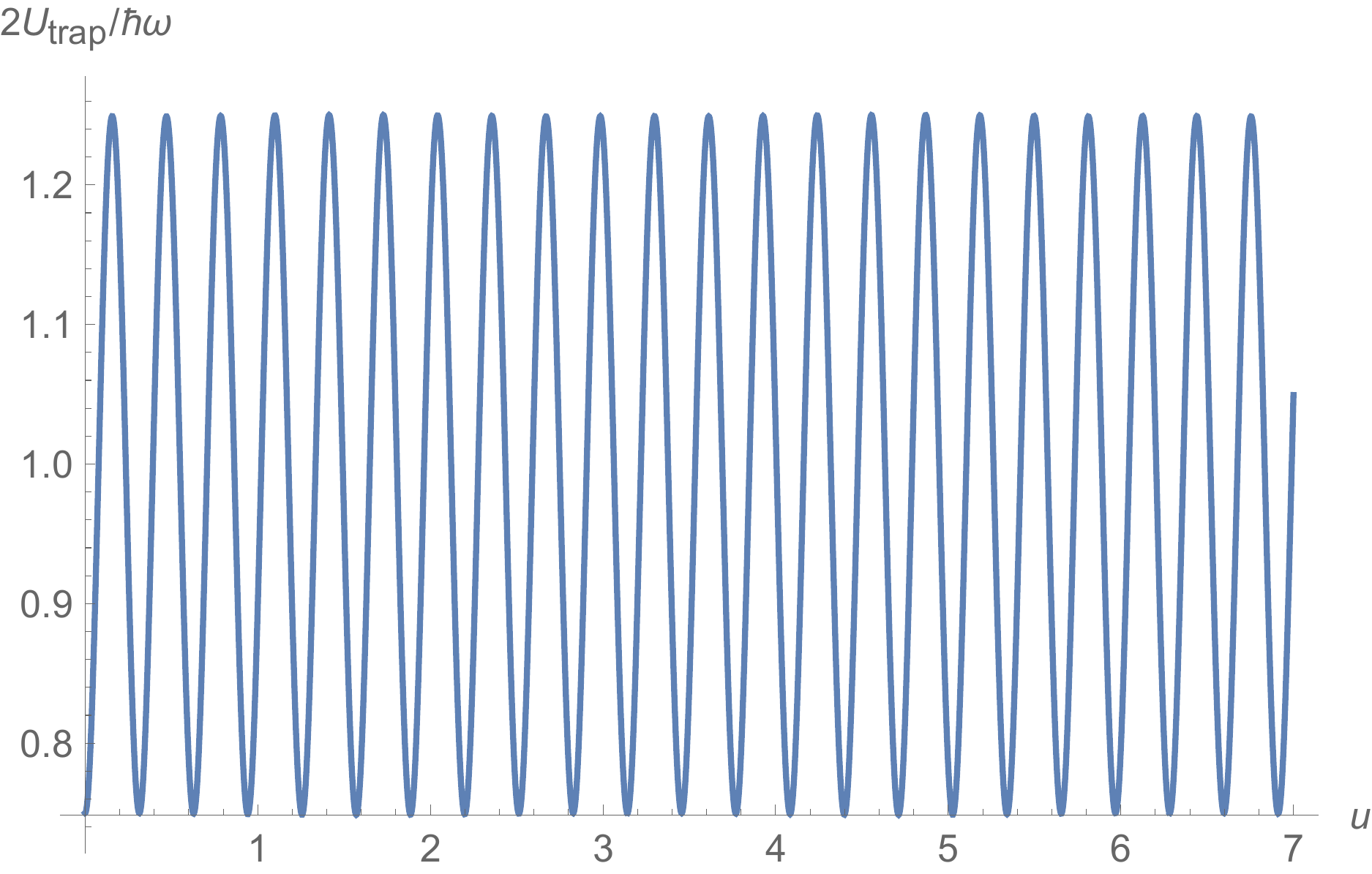}
\caption{The behaviour of normalized energy $2U_{trap}(u)/\hbar\omega$ for values $a=1,\,r=10,\,q=0.25 $. }
\end{figure}
The energy of the trapped particle is given by
\begin{eqnarray}\label{d19}
  U_{trap}(u) &=& \mbox{Tr}(\rho H),\nonumber\\
  &=& \frac{\hbar\omega}{2}\bigg[\frac{\zeta(u)}{2}[a-2q\cos(2ru)]+\frac{1}{2\zeta(u)}+\frac{\alpha^2 (u) (\zeta(u)-1)^2}{2\beta^2(u)\zeta(u)}\bigg],
\end{eqnarray}
and the dimensionless energy $2U_{trap}/\hbar\omega$ is depicted in Fig.2 for the values $a=1,\,r=10,\,q=0.25$. The probability of finding the trapped particle in the nth excited state ($|n\ra$) of a harmonic oscillator ($q=0$), in terms of the normalised time $u$ can be obtained easily as
\begin{eqnarray}\label{pro}
&& P(2n,u)=\frac{1}{\pi\sqrt{\zeta(u)}}\frac{2^{2n}}{(2n)!}\frac{\Gamma^2 (n+1/2)}{\sqrt{\nu^2+\sigma^2}}\bigg(1-\frac{2\nu-1}{\nu^2+\sigma^2}\bigg)^n,\,\,\,n=0,1,2,\cdots,\nonumber\\
&& P(2n-1,u)=0,\,\,\,n=1,2,3,\cdots
\end{eqnarray}
where $\Gamma(\cdot)$ is the gamma function and
\begin{eqnarray}\label{s}
&& \nu=\frac{\zeta(u)+1}{2\zeta(u)},\nonumber\\
&& \sigma=\frac{\alpha(u)(\zeta(u)-1)}{2\beta(u)\zeta(u)}.
\end{eqnarray}
Therefore only even transitions ($2n=0,2,4,\cdots$) are possible. By making use of the identity \cite{Ryzhik}
\begin{equation}\label{ryzhik}
  \sum_{n=0}^{\infty}\frac{\Gamma^2 (n+1/2)(4-4b)^n}{(2n)!}=\frac{\pi}{\sqrt{b}},
\end{equation}
we find
\begin{equation}\label{sump}
\sum_{n=0}^\infty P(2n,u)=1.
\end{equation}
\section{Conclusion}
A novel and simple derivation of quantum propagator of a system described by a general quadratic Lagrangian was presented. The method was based on the explicit form of position and momentum operators in Heisenberg picture and general properties of propagators. From the connection between propagator and density matrix, the corresponding density matrix for a derived quantum harmonic oscillator and a particle trapped in a one dimensional Paul trap was obtained. The total mean energy, work and absorbed heat, Wigner function and transition probabilities were obtained. In the presence of external source ($e(t)\neq 0$), the results obtained for the rate of work done by the oscillator and also the heat absorbed by the oscillator, were different from the usual definitions of these quantities in literature. The method presented here, can be applied to a system consisting of a set of linearly interacting harmonic oscillators straightforwardly.

\end{document}